\pdfoutput=1
\documentclass[a4paper,12pt,english]{article}

\usepackage[pdftex]{graphicx}
\usepackage{graphicx}
\usepackage{colordvi}
\usepackage{fancybox}
\usepackage{cancel}
\usepackage{amsmath}
\usepackage{amssymb}
\usepackage{caption}
\usepackage{appendix}
\usepackage{hyperref} %naredi dokument aktiven (v DVI in PDF)
\usepackage{multirow}
\hypersetup{colorlinks=true, linkcolor=black, urlcolor=blue}

\usepackage{wasysym}

\usepackage{color}
\definecolor{gray}{rgb}{0.5,0.5,0.5}

\newcommand\lsim{\mathrel{\rlap{\lower4pt\hbox{\hskip1pt$\sim$}}
    \raise1pt\hbox{$<$}}}
\newcommand\gsim{\mathrel{\rlap{\lower4pt\hbox{\hskip1pt$\sim$}}
    \raise1pt\hbox{$>$}}}

\newcommand{\beq}{\begin{equation}}
\newcommand{\eeq}{\end{equation}}
\newcommand{\bea}{\begin{eqnarray}}
\newcommand{\eea}{\end{eqnarray}}
\newcommand{\bem}{\begin{pmatrix}}
\newcommand{\eem}{\end{pmatrix}}

\newcommand{\non}{\nonumber}

\begin{document}

\numberwithin{equation}{section}

\begin{flushright}
%MITP-preprint number ?
%December, 2015
\end{flushright}

\bigskip

\begin{center}

{\Large\bf  
Hidden flavor symmetries of SO(10) GUT}
\vspace{1cm}

\centerline{Borut Bajc$^{a,}$\footnote{borut.bajc@ijs.si} and 
Alexei Yu. Smirnov$^{b,c,}$\footnote{smirnov@mpi-hd.mpg.de}}

\vspace{0.5cm}
\centerline{$^{a}$ {\it\small J.\ Stefan Institute, 1000 Ljubljana, Slovenia}}
\centerline{$^{b}$ {\it\small Max-Planck-Institut f\"ur Kernphysik, 
Saupfercheckweg 1, D-69117 Heidelberg, Germany}}
\centerline{$^{c}$ {\it\small International Centre for Theoretical Physics, 
Strada Costiera 11, I-34100 Trieste, Italy}}

\end{center}

\bigskip

\begin{abstract}
The Yukawa interactions of the SO(10)  GUT with fermions in 16-plets (as well as with singlets)
have certain  intrinsic (``built-in'') symmetries which do not depend on the model parameters. 
Thus, the symmetric Yukawa interactions of the 10 and 126 
dimensional Higgses have intrinsic discrete $Z_2\times Z_2$ symmetries, while 
the antisymmetric Yukawa interactions of the 120 dimensional Higgs have a 
continuous SU(2) symmetry. The couplings of SO(10)  singlet fermions  with 
fermionic 16-plets have $U(1)^3$ symmetry.  
We consider a  possibility that  some elements  of 
these intrinsic symmetries are the residual symmetries, which originate from the 
(spontaneous) breaking of a larger symmetry group $G_f$.  
Such an embedding leads to 
the determination of certain elements of the 
relative mixing matrix $U$ between 
the matrices of Yukawa couplings  $Y_{10}$, $Y_{126}$, $Y_{120}$, 
and consequently, to restrictions of 
masses and mixings of quarks and leptons. 
We explore the consequences of such embedding 
using the symmetry group conditions. We show how unitarity emerges from  
group properties and obtain the conditions 
it imposes on the parameters of embedding.  
We find that in some cases the predicted values of elements of $U$  
are compatible with the existing data fits.  
In the supersymmetric version of SO(10) such results are renormalization group 
invariant. 
\end{abstract}

\clearpage

\tableofcontents

\hypersetup{colorlinks=true, linkcolor=red, urlcolor=blue}

\section{Introduction}

In spite of various open questions, Grand Unification \cite{Georgi:1974sy,Pati:1974yy}
is still one of the most appealing 
and motivated scenarios of physics beyond the standard model. The models based on 
SO(10) gauge symmetry \cite{Fritzsch:1974nn,Mohapatra:1979nn,Wilczek:1981iz}
are of special interest since they embed all known fermions 
of a given generation and the right handed neutrinos 
in a single multiplet \footnote{We consider here theories with no extra 
vector-like matter which could mix with SM fermions. This is the case of the majority of  
available models, but may not be the case if SO(10) is coming from $E_6$.}. 
One of the open questions is to understand  
the flavor structures - observed fermion masses and mixings,  
which SO(10) unification alone can not fully address\footnote{
Partially it sometimes can: for example, $b-\tau$ unification can be related 
to the large atmospheric mixing angle in models with dominant type II seesaw \cite{Bajc:2002iw}.}. 
Moreover, embedding of all the fermions in a single multiplet looks at odds with  
different mass hierarchies and mixings, and in particular with the strong difference of 
mixing patterns of quarks and leptons.  

The Yukawa sector of the renormalizable\footnote{To realize eventually our scenario these 
Yukawa couplings  should be VEVs of fields which transform non-trivially under some flavor 
group $G_f$ - so they will be non-renormalizable, 
or we should ascribe charges to the Higgs multiplets.} 
version of SO(10) GUT \cite{renSO10GUT} with three generations of matter fields in $16_F$ is given by 
\beq
\label{WYukawa}
{\cal L}_{Yukawa}=16_F^T\left(Y_{10} 10_H + Y_{126} \overline{126}_H  
+ Y_{120} 120_H  \right)16_F, 
\eeq
where the $3\times3$ matrices  of Yukawa couplings, $Y_{10}$, $Y_{126}$ and $Y_{120}$ 
correspond to  Higgses in $10_H$, $\overline{126}_H$ and $120_H$. 
The masses and mixings of the Standard Model (SM) 
fermions are determined by these  Yukawa couplings $Y_{a}$, 
the Clebsch-Gordan coefficients and the VEV's of the light Higgs(es). 
So, to make predictions for the masses and mixing one needs,  
in turn, to  determine the matrices $Y_{a}$  ($a = 10, 126, 120$). 

There are various attempts 
to impose a flavor symmetry on the Yukawa interaction (\ref{WYukawa}) 
to restrict the mass and mixing parameters,  
see for example \cite{continuous} for continuous symmetries, 
\cite{discrete,Ferreira:2015jpa,Ivanov:2015xss} for discrete symmetries, and 
\cite{reviews}  for reviews. In most cases flavor symmetries appear 
as horizontal symmetries - which are independent 
of the vertical gauge symmetry SO(10). \\

Two interesting ideas have been discussed recently 
which employ  an interplay between the GUT symmetry and flavor symmetries and 
may lead to deep relation between them.\\  

1. Existence of ``natural'' (``built-in") or intrinsic  flavor symmetries \cite{Lam:2014kga}. 
Examples are  known from the past that some approximate flavor symmetries 
can arise from the ``vertical'' gauge symmetries. 
One of these is the antisymmetry of the Yukawa couplings of the lepton 
doublets with charged scalar singlet. The neutrino mass matrix generated at 
1-loop (Zee model \cite{Zeemodel}) has specific flavor structure  
with zero diagonal terms. 

It is well known that SO(10) have such flavor symmetries. 
The three terms in (\ref{WYukawa}) have  symmetries dictated 
by ``vertical"  SO(10): symmetricity of the Yukawa 
coupling matrices of the 10-plet and 126-plet of  
Higgses and antisymmetricity of the Yukawa coupling  
matrix of the 120-dimensional Higgs multiplets:  
\beq
Y_{10,126}^T = Y_{10,126}\;,\;\;\;Y_{120}^T = - Y_{120}. 
\eeq
The first equality (symmetricity) implies a $Z_2 \times Z_2$ symmetry \cite{Lam:2014kga}. 
For the antisymmetric matrix (second equality) the symmetry $(Z_2)$ has been taken in 
\cite{Lam:2014kga} (or $(Z_2)^2$ if negative determinants are allowed).\\ 

2. Identification of the natural symmetries with residuals 
of the flavor symmetry \cite{Lam:2014kga}. 
This idea is taken from the residual symmetry approach developed to explain 
the lepton mixing. It states  that   some or all elements of the  
natural symmetries of SO(10) are actually the residual symmetries which 
originate from the breaking of a bigger flavor symmetry group 
$G_f$  \cite{Lam:2007qc,Lam:2008sh,Lam:2009hn,Grimus:2009pg,Grimus:2008vg}. 
In \cite{Lam:2014kga} it was proposed to embed the residual $(Z_2)^n$,  
which are reflection symmetries, 
into the minimal group with a three-dimensional 
representation. This leads to the Coxeter group and  finite Coxeter groups of rank 3 and 4 
have been considered. 
The embedding of natural symmetries into the flavor (Coxeter) group 
imposes restrictions on the 
structure of $Y_a$ and consequently on the mass matrices,  
which reduces the number of free parameters.\\

In this paper we further elaborate on realizations of these ideas, although   
from a different point of view. 
While the intrinsic symmetries of $Y_{10}$ and $Y_{126}$ are $Z_2 \times Z_2$, 
as in \cite{Lam:2014kga}, we find that $Y_{120}$ has a bigger symmetry - SU(2).  
Furthermore,  we consider  the situation when SO(10) singlet fermions are 
present. From the embedding of intrinsic symmetries and with the 
use of symmetry group relations \cite{Hernandez:2012ra,Hernandez:2012sk}
we obtain predictions for the elements of the {\it relative mixing matrix} $U_{a-b}$ 
($a, b = 10, 126, 120$)  between the Yukawa couplings $Y_a$ and $Y_b$ 
($U_{a-b}$  connects the bases in which matrices $Y_a$ and $Y_{b}$ are diagonal). 
These unitary matrices $U_{a-b}$ are basis independent, in contrast to the matrices 
$Y_a$ and $Y_b$ themselves.  We re-derive these relations and elaborate 
on the unitarity condition, showing that it follows from group properties.  
We confront the predictions with the results of some available data fits. 

The paper is organized as follows. In sect.  
\ref{intrinsic} we explore the intrinsic symmetries of the 
SO(10) Yukawa couplings. 
In sect. \ref{embb} we identify (part of) the intrinsic symmetries with the residual symmetries 
and consider their embedding into a bigger flavor group. 
Using the symmetry group relations we obtain  predictions for 
different elements of the relative  matrix $U$. 
We elaborate on the unitarity condition which gives additional bounds 
on the parameters of embedding. 
We consider separately the embeddings of the $120_H$ couplings. This case has 
not been covered in \cite{Lam:2014kga} and we develop various methods to deal with it. 
In sect. \ref{data} we confront our predictions for the mixing 
matrix elements with the results obtained from existing fits of data. 
In sect. \ref{singlets} we consider  symmetries in the presence of the 
SO(10) fermionic singlets. In sect. \ref{intrinsicmixing} we summarize the 
concept of intrinsic symmetry and the relative mixing matrix. 
Summary of our results and conclusion are presented in sect. \ref{outlook}. 
We compare our approach with that in \cite{Lam:2014kga} 
in Appendix \ref{Lam}, suggesting an equivalence. 

\section{\label{intrinsic}Intrinsic flavor symmetries of SO(10)}
%%%%%%%%%%%%%%%%ssss2%%%%%%%%%%%%%%%%%%%%%%%%%%%%%%%%%%%%%%%%

\subsection{Relative mixing matrices}
%%%%%%%%%%%%%%%%%%%%%%%%%%%%%%%%%%%%%%%%%%%%%%%%%%%%%%%%%%%%%%%%%%%%%

The  matrices of Yukawa couplings  are basis dependent.  
It is their eigenvalues and the relative mixings which have physical meaning. 
The relative mixing matrices, which are the main object of this paper,   
are defined in the following way. The symmetric matrices $Y_{10}$ and 
$Y_{126}$  can be  diagonalized with 
the unitary transformation matrices $U_{10}$  and $U_{126}$ as  
\beq
Y_{10}= U_{10}^* Y_{10}^d U_{10}^\dagger ,
\label{10diag}
\eeq
and 
\beq
Y_{126}= U_{126}^*Y_{126}^dU_{126}^\dagger.
\label{u126}
\eeq
Mixing is generated if the matrices $Y_a$ 
can not be diagonalized simultaneously. 
The relative mixing matrix $U_{10-126}$ is given by
\beq
\label{U10126}
U_{10-126} = U_{10}^\dagger U_{126}.     
\eeq
This  matrix, in contrast to matrices of Yukawa couplings,  does not depend on basis 
and  has immediate physical  meaning. In a sense, it is the analogy of the PMNS (or CKM) matrix which 
connects bases  of mass states of neutrinos and charged leptons. 
Similarly we can introduce the relative mixing matrices for other Yukawa 
coupling matrices as 
\beq
U_{a-b}=U_a^\dagger U_b, 
\eeq
e.g.,  $U_{10-120}$, $U_{120-126}$, {\it etc}. 

The symmetry formalism we present below (symmetry group relations)  will determine 
elements of the relative matrices immediately without consideration of the 
symmetric matrices $Y_a$ and their diagonalization.

%%In \cite{} they  are called natural symmetries [[with a bit different applications... ]]

\subsection{Intrinsic symmetries}
%%%%%%%%%%%%%%%%%%%%%%%%%%%%%%%%%%%%%%%%%%%%%%%%%%%%%%%%%%%%%%%

All the terms of the Lagrangian (\ref{WYukawa})
have the same fermionic 
structure, being the Majorana type bilinears of $16_F$.  
This by itself implies certain symmetry. 
For definiteness let us  consider the basis of three $16_F$ plets 
in which the Yukawa coupling of the $10$-plet is diagonal: 
\beq
Y_{10}= Y_{10}^d.  
\eeq
In this basis the Yukawa matrix of $\overline{126}_H$ 
(being in general non-diagonal) 
can be diagonalized  by  the unitary matrix $U_{126}$ as in  (\ref{u126}).  
In this basis $U_{126}$ gives immediately the relative mixing matrix 
$U_{10 - 126}  =  U_{126}$. 
It is straightforward to check that 
the symmetric matrices $Y_{10}^d$ and $Y_{126}$  
are invariant with respect to transformations  
\bea
{S_j^d} Y_{10}^dS_j^d = Y_{10}^d &,&j=1,2,3, \\
(S_{126})_i^T~ Y_{126}~ (S_{126})_i = Y_{126}&, &i=1,2,3, 
\eea
where 
\beq
(S_{126})_i  = U_{126} S_i^d U_{126}^\dagger,  
\eeq
and the diagonal transformations equal 
\beq
S_1^d = {\rm diag} (1,~ - 1,~ - 1), ~~~~~ S_3^d = {\rm diag} (- 1,~ - 1,~ 1),   
\label{eq:reflect}
\eeq
$S_2^d = S_1^d S_3^d$.   
(We use generators with  ${\rm Det} [S_i] = +1$,   
so that they can form  a subgroup of SU(3)).  

The transformations (\ref{eq:reflect}) can be written as 
\beq
\label{sid}
\left(S_j^d\right)_{ab} = 2\delta_{aj}\delta_{bj}-\delta_{ab}, 
\eeq
and $a,b=1,2,3$.  All these  transformations  (reflections) obey 
\beq
(S_j)^2 = (S_j^d)^2 =  \mathbb{I}. 
\eeq
Thus,  $Y_{10}^d$ is invariant under the group of transformations 
 $G_{10} =  Z_2\times Z_2$  consisting of elements 
\beq
\label{Z2Z210}
G_{10}=  \{1,S_1^d,S_2^d,S_3^d\}.  
\eeq
The matrix $Y_{126}$ is invariant under another, 
$G_{126} =  Z_2\times Z_2$ group consisting of $U-$ transformed 
elements 
\beq
G_{126}=
U_{126}\{1,S_1^d,S_2^d,S_3^d\}U_{126}^\dagger,  
\eeq
where $U_{126}$ is defined in (\ref{u126}). 

This intrinsic symmetry  is always present independently 
of  parameters of the model due to the symmetric Yukawa matrices $Y_{10}$ 
and $Y_{126}$ \cite{Lam:2014kga} which follow from SO(10) symmetry. \\ 

In the case of antisymmetric Yukawa interactions  
of $120_H$ the situation is  different. 
The antisymmetric matrix $Y_{120}$ can be put in the canonical form 
\beq
\label{Y120}
Y_{120}^c=
\bem
0 & 0 & 0 \cr
0 & 0 & x \cr
0 & -x & 0 
\eem
\eeq
by the unitary transformation $U_{120}$ as
\beq
Y_{120}=U_{120}^*Y_{120}^cU_{120}^\dagger. 
\eeq
The matrix (\ref{Y120}) is invariant with respect to SU(2)$\times$U(1) transformations  
\beq
g^T Y_{120}^cg=Y_{120}^c. 
\eeq
Again we will bound ourselves to group elements with ${\rm Det}(g)=1$, 
keeping in mind possible embedding into $SU(3)$. Then 
there is no U(1), and therefore 
\beq
G_{120} = SU(2).  
\eeq
The $SU(2)$ transformation element $g$  can be written as
\beq
g(\vec{\phi})=
\bem
1 & 0 \cr
0 & \exp{\left(i\vec{\phi}\vec{\tau}\right)}
\eem=
\bem
1 & 0 \cr 
0 & \cos{\phi}+i\frac{\vec{\phi}\vec{\tau}}{\phi}\sin{\phi}
\eem
\label{eq:g120}
\eeq
with $\vec{\phi}=(\phi_1,\phi_2,\phi_3)$, 
$\phi\equiv|\vec{\phi}|\in[0,\pi]$ and $\vec{\tau}$ being the Pauli matrices.  

Although the symmetry of the Yukawa matrix connected 
to the 120-plet is continuous, we should use only its  discrete subgroup to be 
a part of  $G_f$, since $G_f$ itself has been assumed  
to be discrete. This means that the angle $\vec{\phi}$ should take 
discrete values such that 
\beq
\left(g(\vec\phi)\right)^p=\mathbb{I}
\eeq
for some integer $p$. The angle can be parametrized as   
\beq
\vec{\phi}=2\pi\frac{n}{p}\hat{\phi}\;\;\;,\;\;\;n = 1,\ldots, p-1,
\eeq
where 
$\hat{\phi} \equiv \vec{\phi}/|\vec{\phi}|$ (so that $\hat{\phi}^2=1$). 
In this paper we will consider a $Z_p$ subgroup of the Abelian 
$U(1)\subset SU(2)$. So, 
the elements $g_\phi,g_\phi^2,\ldots,g_\phi^{p-1}$ can be written as 
\beq
\label{gdiscrete}
g_\phi^n  =
\bem
1 & 0 \cr
0 & \exp{\left(i2\pi (\vec{\tau}\hat{\phi})n/p\right)}  
\eem. 
\eeq

More on intrinsic symmetries and the mixing matrices can be found in 
Appendix \ref{intrinsicmixing}. Intrinsic symmetries for the  SO(10) singlets 
are discussed in sect. \ref{singlets}. 
 
We assume throughout this paper that the Higgs multiplets are uncharged with respect to $G_f$. 
Introduction of Higgs charges can lead to suppression of some Yukawa couplings but 
does not produce  the flavor structure of individual interactions. 

\section{\label{embb}Embedding intrinsic symmetries}
%%%%%%%%%%%%%%%%%%%%%%%%%%%%%%%%%%%%%%%%%%%%%%%%%%%%%%%%%%%%%%%%%
Following \cite{Lam:2014kga} we assume that the intrinsic symmetries formulated in the previous 
section are actually residual 

which result from the breaking of a larger (flavor) symmetry group $G_f$. 
In other words, some of the symmetries $G_{10}$ and $G_{126}$ or/and $G_{120}$ are embedded 
into $G_f$. In the following we will derive various constraints on the relative mixing matrix 
$U$ between two Yukawa matrices. 

\subsection{\label{ressym}Embedding of two transformations}
%%%%%%%%%%%%%%%%%%ssss3%%%%%%%%%%%%%%%%%%%%%%%%%%%%%%%%%%%%%%%%%%%%

We recall the symmetry group relation formalism  \cite{Hernandez:2012ra,Hernandez:2012sk} 
adopted to our $SO(10)$ case. The formalism  
allows to  determine (basis independent) elements  of the relative mixing matrix 
immediately without explicit construction of  Yukawa matrices. 
Let us first consider the Yukawa couplings of $10_H$ and  $126_H$. 
Suppose the covering group $G_f$ contain $S_j^d\in G_{10}$ and $S_i\in G_{126}$. 
Since $S_i, S_j^d\in G_{f}$, 
the product $S_i S_j^d $ should also belong to $G_f$: 
$S_i S_j^d\in G_{f}$. Then the condition of finiteness of $G_f$ requires  
that a positive integer $p_{ji}$ exists such that 
\beq
\label{s1s3d}
\left(S_i S_j^d\right)^{p_{ji}}   =\mathbb{I}. 
\eeq
This is the symmetry group relation \cite{Hernandez:2012ra,Hernandez:2012sk} 
which we will use in our further study. 
Inserting $S_i =  US_i^d U^\dagger$\footnote{In this and the next section $U\equiv U_{10-126}$.} 
into (\ref{s1s3d}) we obtain
\cite{Hernandez:2012ra,Hernandez:2012sk}
\beq
\label{Wijpje1}
(W_{ij})^{p_{ji}}=\mathbb{I}, 
\eeq
where 
\beq
\label{Wij}
W_{ij} \equiv US_i^dU^\dagger S_j^d. 
\eeq
Furthermore,  we will impose the condition 
\beq
{\rm Det} [W_{ij}] = 1
\eeq
keeping in mind a possible embedding into $SU(3)$. 
We will comment on the case of negative determinant later. \\

%%\subsubsection{$Z_2^{(10)}\times Z_2^{(126)}$}
%%%%%%%%%%%%%%%%%%%%%%%%%%%%%%%%%%%%%%%%%%%%%%%%%%%%%%%%%%

The simplest possibility is the residual symmetries 
$Z_2^{(10)} \times Z_2^{(126)}$,  that is 
$Z_2$ for $Y_{10}$ and  another $Z_2$ for $Y_{126}$. 
In this case  
the flavor symmetry group $G_f$ is always a finite von Dyck group (2, 2, p), since 
\beq
\frac{1}{2}+\frac{1}{2}+\frac{1}{p}>1
\eeq
for any positive integer $p$.

Let us elaborate on the constraint   
(\ref{s1s3d}) further, providing
derivation of the relations slightly different  to that in
\cite{Hernandez:2012ra,Hernandez:2012sk}. 
According to the Schur decomposition we can present $W_{ij}$ in the form  
\beq
\label{WSchur}
W_{ij}=V W_{ij}^{upper}V^\dagger, 
\eeq
where $V$ is a unitary matrix and $W_{ij}^{upper}$ is 
an upper triangular matrix, the so called Schur form of $W_{ij}$. 
Since unitary transformations do not change the trace,  we have from 
(\ref{WSchur}) 
\beq
{\rm Tr}[W_{ij}]  = {\rm Tr}[W_{ij}^{upper}]. 
\eeq
The diagonal elements of $W_{ij}^{upper}$ are the (in general complex) 
eigenvalues of $W_{ij}$ which we  denote by $\lambda_\alpha$. Therefore,   
\beq
{\rm Tr}[W_{ij}^{upper}] = a_{p_{ji}},    
\label{trace1}
\eeq
where 
\beq
a_{p_{ji}} \equiv  \sum_\alpha \lambda_\alpha.
\label{trace}
\eeq
Inserting (\ref{WSchur}) into condition 
(\ref{Wijpje1}) and using unitarity of $V$ we obtain
\beq
(W_{ij}^{upper})^{p_{ji}} = 
diag(\lambda_1^{p_{ji}}, \lambda_2^{p_{ji}}, \lambda_3^{p_{ji}})  = 
\mathbb{I}
\eeq
the off-diagonal elements in the LH side should be zero to match with the RH side). 
Consequently,  the eigenvalues of $W$ equal  the $p_{ji}$ - roots of unity: 
\beq
\lambda_\alpha = ~ ^{p_{ji}}\sqrt{1}. 
\eeq
Finally, Eq. (\ref{trace1}) gives  
\beq
{\rm Tr}[W_{ij}] = a_{p_{ji}}, 
\label{eq:trace}
\eeq
where $a_{p_{ji}}$ is defined in (\ref{trace}). 

The  $p_{ji}-$roots of unity can be parametrized as 
\beq 
\lambda = \exp{\left(i2\pi k_{ji}/p_{ji}\right)}\;\;\;,\;\;\;k_{ji}=1,\ldots,p_{ji}-1. 
\label{eigenval}
\eeq
For $p \geq 3$ the number of p-roots is larger than 3,  and therefore there is an ambiguity 
in selecting the three values to compose $a_{p_{ji}}$. 
However,  not all combinations can be used, and certain restrictions 
will be discussed in the following.  

Restriction on $a_{p_{ji}}$ arises from the following consideration. 
The eigenvalues  $\lambda_\alpha$ satisfy the characteristic
polynomial equation  $W_{ij}$:
\beq
\label{charpol}
{\rm Det} \left(\lambda \mathbb{I}-W_{ij}\right)= \lambda^3 - a_{p_{ji}}\lambda^2+
a_{p_{ji}}^*\lambda - 1 = 0, 
\eeq
where $a_{p_{ji}}$ is defined in  (\ref{trace})\footnote{
This can be obtained noticing that
\beq
{\rm Det}\left(\lambda \mathbb{I} - W_{ij}\right) =
(\lambda - \lambda_1)(\lambda - \lambda_2)(\lambda - \lambda_3),
\eeq
$\left|\lambda_i\right|^2=1$ and ${\rm Det} (W_{ij}) \equiv  \lambda_1 \lambda_2 \lambda_3  = 1$.
}.
Consider the conjugate of Eq. (\ref {charpol}). 
Using the expression for (\ref{Wij}) and taking into account that 
$\left(S_i^d\right)^2=\mathbb{I}$ we obtain 
\beq
W_{ij}^\dagger =  S_j^dW_{ij}{S_j^d}.  
\eeq
This in turn gives for the  LHS of the conjugate equation 
\beq
{\rm Det}\left(\lambda^* \mathbb{I} - W_{ij}^\dagger \right) = 
{\rm Det}\left(\lambda^* \mathbb{I} - S_j^dW_{ij}{S_j^d} \right) = 
{\rm Det}\left[S_j^d ( \lambda^* \mathbb{I} - W_{ij}){S_j^d} \right] = 
{\rm Det}\left(\lambda^* \mathbb{I} - W_{ij} \right). 
\eeq
Therefore the set of eigenvalues  $\{\lambda_\alpha\}$ 
coincides with the set  $\{\lambda_\alpha^*\}$ \cite{Fonseca:2014koa}.  
Then it is easy to check that this is possible only if one of $\lambda_\alpha$ 
equals unit,  e.g. $\lambda_1 = 1$, 
and two others are conjugate of each other: $\lambda_3 =  \lambda_2^* \equiv \lambda$. 
Thus, 
\beq
\label{ap}
a_{p_{ji}} = a_{p_{ji}}^* = 1 + \lambda + \lambda^*  = 1 + 2 {\rm Re} \lambda, 
\eeq
or explicitly,   
\beq
a_{p_{ji}} = 1+2\cos{\left(2\pi k_{ji}/p_{ji}\right)}
= -1 + 4 \cos^2{\left(\pi k_{ji}/p_{ji}\right)}. 
\label{eq:for-a}
\eeq

On the other hand, from definitions of $S_j$ (\ref{sid}) and (\ref{Wij}), 
we find explicitly 
\beq
\label{apUji}
{\rm Tr} \left(W_{ij}\right) =    4\left|U_{ji}\right|^2 - 1  
\eeq
or using (\ref{eq:trace}) (see also \cite{Esmaili:2015pna}) 
\beq
\left|U_{ji}\right|^2 = \frac{1}{4} \left(1 - a_{p_{ji}} \right). 
\label{eq:uvera}
\eeq
Notice that the trace (\ref{apUji}) is real, and therefore 
$a_{p_{ji}} = a_{p_{ji}}^*$, leading to the form (\ref{ap}). 
Finally, inserting  $a_{p_{ji}}$ from (\ref{eq:for-a}) we obtain 
\beq
\label{Uji}
\left|U_{ji}\right|=\left|\cos{\left(\pi k_{ji}/p_{ji}\right)}\right|. 
\eeq
Similar expression has been obtained before 
in \cite{Blum:2007jz} 
in the Dihedral group model for the Cabibbo angle ($V_{us}$). 
The expression appears also in \cite{Byakti:2016rru}. 

Thus, we obtain thus a relation for a single element of the matrix $U$, 
as  the consequence of the $Z_2^{(10)}\times Z_2^{(126)}$ residual symmetry. 
The element $\left|U_{ji}\right|$   is determined by two 
discrete parameters -- arbitrary integers 
$p_{ji}$ and $k_{ji} = 0,\ldots,p_{ji} -1$. 
The expression does not depend on the selected $S_i$. The elements $S_i$  and $S_j$ just 
fix the $ij-$ element of the matrix $U$, but not its value, 
the value is determined by $p_{ji}$ and $k_{ji}$. 

Allowing also ${\rm Det} (W_{ij}) = -1$ we generalize (\ref{Z2Z210}) into 
\beq
\left(Z_2\times Z_2\right)_{10}\to\{1,S_1^d,S_2^d,S_3^d\}\cup\{-1,-S_1^d,-S_2^d,-S_3^d\},
\eeq
while in (\ref{Wij}) $S_i^d$ (and/or $S_j^d$) can be replaced by $-S_i^d$ (and/or $-S_j^d$). 
A difference from the previous case comes only if in $W_{ij}$ the two diagonal 
group elements have opposite signs of determinants. 
In this case we have  
${\rm Det} (W_{ij}) =  \lambda_1 \lambda_2 \lambda_3  = -1$ 
and since now
one of the eigenvalues needs to be $\lambda_1 = -1$\footnote{
The eigenvalues $\lambda_{1,2,3}$ of $W_{ij}$ satisfy ($\left|\lambda_\alpha\right|^2=1$)
\bea
0= {\rm Det}\left(\lambda \mathbb{I}-W_{ij}\right)
&=&
\lambda^3-\left(\lambda_1+\lambda_2+\lambda_3\right)\lambda^2
+\left(\lambda_1\lambda_2+\lambda_2\lambda_3+\lambda_3\lambda_1\right)\lambda-
\lambda_1\lambda_2\lambda_3\non\\
&=&\lambda^3-a_{p_{ji}}\lambda^2 + {\rm Det} \left(W_{ij}\right)a_{p_{ji}}^*\lambda-
{\rm Det} \left(W_{ij}\right). 
\eea
If $a_{p_{ji}}^*=a_{p_{ji}}$, 
one eigenvalue is equal to ${\rm Det} (W_{ij})$.
}, 
we obtain that $\lambda_2 \lambda_3  =  1$ 
or $\lambda_2 =   \lambda_3^* \equiv \lambda$.
Then 
$a_{p_{ji}} = -1 + \lambda + \lambda^*  = -1 + 2 Re (\lambda)$, 
and consequently, 
\beq
\label{Ujiminus}
\left|U_{ji}\right|=\left|\sin{\left(\pi k_{ji}/p_{ji}\right)}\right|,  
\eeq
(as compared with (\ref{Uji})). 

\subsection{\label{unitarity}Embedding of bigger residual symmetries and  Unitarity}
%%%%%%%%%%%%%%%%%%%%%%%%%%%%%%%%%%%%%%%%%%%%%%%%%%%

Following the derivations in \cite{Fonseca:2014koa,Fonseca:2015kya} we summarize here 
the embedding of bigger residual symmetries, when 
we take  $Z_2 \times Z_2$  
from one of the interactions ($10_H$ or $126_H$) 
and one  $Z_2$ from the other interaction. 
Now there are three generating elements: 
for $Z_2^{(10)} \times Z_2^{(10)} \times Z_2^{(126)}$
the matrix $Y_{10}$ is invariant under   
$S_j^d$ and $S_k^d$ ($j \neq k$),  whereas $Y_{126}$ --   
under  $S_i$.  
Consequently,  we have two symmetry group conditions: 
\beq
\label{WijWik}
(U S_i^dU^\dagger S_j^d)^{p_{ji}} =\mathbb{I}, ~~~~ 
(U S_i^dU^\dagger S_k^d)^{p_{ki}} =\mathbb{I} 
\eeq
which determine two elements of the matrix $U$ from the same column $i$:    
$|U_{ji}|$ and $|U_{ki}|$. Repeating the same procedure of the previous section we obtain   
\beq
\label{Ujiplus}
\left|U_{ji}\right|=\left|\cos{\left(\pi k_{ji}/p_{ji}\right)}\right|, ~~~~
\left|U_{ki}\right|=\left|\cos{\left(\pi k_{ki}/p_{ki}\right)}\right|. 
\eeq

The second possibility is   
$Z_2^{(10)} \times Z_2^{(126)}  \times Z_2^{(126)}$  
with one generating element for $Y_{10}$ and two for $Y_{126}$.  
This gives also two symmetry group conditions but for two elements 
in the same row of $U$. This is  enough to determine the whole 
row (or column in the first case)  from unitarity. 
Possible values of matrix elements for this case have 
been classified in whole generality \cite{Fonseca:2014koa,Fonseca:2015kya}.

Using the complete symmetry 
$Z_2^{(10)}  \times Z_2^{(10)} \times  Z_2^{(126)}  \times Z_2^{(126)}$ 
one can fix 4 elements of $U$, and consequently, due to unitarity, 
the whole matrix  $U$. This matrix is necessarily of the type classified in 
\cite{Fonseca:2014koa,Fonseca:2015kya}.

Notice that values of the elements of the relative matrix $U$ 
have been obtained using different 
group elements $S_j$ (for fixed $S_i$)  essentially independently. 
They were determined by 
the independent parameters $p_j$, $k_j$. However, there are  
relations between the group elements $S_j$  which, 
as we will see, lead  to relations between parameters $p_j$, $k_j$, 
which are equivalent to relations required by unitarity of the matrix $U$. 

According to (\ref{Uji}) $|U_{ji}| \leq 1$ 
for any pair of values of $k$ and $p$. 
For two elements in the same line or column unitarity 
requires  
\beq
\cos^2{\left(\pi k_1/p_1\right)}+\cos^2{\left(\pi k_2/p_2\right)} \leq  1 
\label{unit-ineq}
\eeq
and it is not fulfilled  automatically. 
(In this section we omit the second index of  $k$  and $p$, 
which is the same for both. Keeping in mind that both are 
from the same line or the same column.)  
Furthermore, the  inequality (\ref{unit-ineq})  can not be satisfied for arbitrary 
$k_i$ and  $p_i$, and therefore 
gives certain bounds on these parameters. This, in turn,  affects the embedding 
(covering group).  
In what follows we will consider such restrictions on parameters 
$k$  and $p$ that follow from relations between the group elements. \\
 
%\subsection{\label{unitarity}Consistency condition and unitarity}
%%%%%%%%%%%%%%%%%%%%%%%%%%%%%%%%%%%%%%%%%%%%%%%%%%%%%%%%%%%%%%%%%

The elements of  $Z_2 \times Z_2$  
group in 3 dimensional representation  (\ref{eq:reflect}) or (\ref{sid})
satisfy the following equalities 
\beq
\sum_{i=1}^3 S_i^d = -\mathbb{I}, 
\label{sumsi}
\eeq
and 
\beq
{\it Tr} \left(S_i\right) = -1, ~~~~i = 1, 2, 3. 
\label{trsi}
\eeq
Let us find the corresponding relations between the parameters 
$p_i$ and $k_i$. 
Summation over the index $i$ of the traces ${\rm Tr} [W_{ij}]$, 
where $W_{ij}$ is given in eq. (\ref{Wij}),  gives 
\beq
\sum_i {\rm Tr} [W_{ij}] = {\rm Tr} \left[\sum_i  W_{ij}\right] = 
{\rm Tr} \left[U \left(\sum_i S_i^d \right) U^{\dagger}S_j^d  \right].  
\eeq
The last expression in this formula together with equalities 
(\ref{sumsi}) and (\ref{trsi})   gives $- Tr [S_j^d] = 1$.   
Therefore $\sum_i {\rm Tr} [W_{ij}] = 1$ and  according 
to (\ref{eq:trace}) we find 
\beq
a_{p_1}+ a_{p_2}+ a_{p_3} = 1. 
\label{eq:sum-a}
\eeq
Finally, insertion of  expressions 
for $a_{p_i}$ in eq. (\ref{eq:for-a}) leads to 
\beq
\cos^2{\left(\pi k_1/p_1\right)} + \cos^2{\left(\pi k_2/p_2\right)} 
+ \cos^2{\left(\pi k_3/p_3\right)} = 1. 
\label{cos-rel}
\eeq
This  coincides with the unitarity condition: Eq. (\ref{cos-rel}) 
is nothing but 
$\sum_i |U_{ij}|^2 = 1$,   where the elements are expressed via 
cosines (\ref{Uji}). 
So,  the unitarity condition  is encoded in the relation (\ref{sumsi}) 
which is  equivalent to the unitarity.  
Thus,  the unitarity condition which imposes relations between  
$p_{j}$ and $k_{j}$ can be obtained automatically 
from properties of the group elements. 

The condition is highly non-trivial since it should be satisfied for 
integer values of $p_i$ and $k_i$. 
It can be fulfilled for specific choices of $(k_1/p_1,k_2/p_2,k_3/p_3)$. 
There are just few cases which can satisfy  
(\ref{cos-rel}).  
Some of these constraints  have been found in 
\cite{Feruglio:2012cw,Hagedorn:2013nra}  from specific assumptions on $G_f$.  
In general, it has been shown 
\cite{Fonseca:2014koa,Fonseca:2015kya} (see also \cite{Byakti:2016rru}) that  the only possibilities are 
\beq
 \{c_i \} \equiv (c_1,~ c_2,~ c_3) = 
\left(\frac{1}{\sqrt{2}},~\frac{1}{2},~\frac{1}{2}\right), 
\label{ccc1}
\eeq
where 
\beq
c_i \equiv \cos \left(\pi \frac{k_i}{p_i}\right), ~~~~ i = 1, 2, 3. 
\eeq
The values in (\ref{ccc1})  correspond to  
\beq
(k_1/p_1,~ k_2/p_2, ~ k_3/p_3) = \left(\frac{1}{4},~\frac{1}{3},~\frac{1}{3}\right). 
\eeq
Another solution, 
\beq
\{c_i \} 
= \left(\frac{1}{2},\frac{\phi}{2},\frac{1}{2\phi}\right), 
\label{ccc2}
\eeq  
where 
\beq
\phi = \frac{\sqrt{5}+1}{2} 
\eeq
is the golden ratio. In this case   
\beq
(k_1/p_1,~ k_2/p_2, ~ k_3/p_3) = \left(\frac{1}{3},~\frac{1}{5},~\frac{2}{5}\right)
\eeq
Finally, 
\beq 
\{c_i \} =  \left(\cos{\alpha},~\sin{\alpha},~ 0\right)
\label{ccc3}
\eeq
with 
\beq
\alpha =  \pi k_0/p_0\;\;\;,\;\;\;1\leq k_0\leq p_0/2\;\;\;,\;\;\;k_0\in\mathbb{Z}.
\eeq
They  correspond to
\beq
(k_1/p_1,~ k_2/p_2, ~ k_3/p_3) = 
\left(\frac{k_0}{p_0},~\frac{1}{2}- \frac{k_0}{p_0},~\frac{1}{2}\right). 
\eeq
For instance for  $k_0/p_0 = 1/2$ we obtain  ${c_i} =  (0, 1, 0)$, 
for $k_0/p_0 = 1/3$:  ${c_i} =  (1/2, \sqrt{3}/2, 0)$, 
for $k_0/p_0 = 1/4$:  ${c_i} =  (1/ \sqrt{2}, 1/\sqrt{2}, 0)$,  {\it etc.}

\subsection{The case with   $120_H$ coupling}
%%%%%%%%%%%%%%%%%%%%%%%%%%%%%%%%%%%%%%%%%%%%%%%%%%%%%%%%%%%%

The symmetry transformation of $Y_{120}$ is given by the elements $g_\phi^n$ of a discrete 
subgroup $Z_p$ of $U(1) \subset SU(2)$ (\ref{gdiscrete}). 
Since $g_\phi^2\ne\mathbb{I}$ for $p>2$, the embedding symmetry group is 
not a Coxeter group, and so this analysis goes beyond the assumptions of \cite{Lam:2014kga}. 
If we assume that the element $g_\phi$ from the $Z_p$ intrinsic symmetry of $Y_{120}$ and 
the element $S_j^d$ from the $Z_2$ intrinsic symmetry of $Y_{10}$ (or $Y_{126}$) are residual 
symmetries, from the definition of a group this is true also for all $g_\phi^n$, $n=1\ldots,p-1$. 
Therefore, the symmetry group relations  
now contain the products of $U g_{\phi}^n U^\dagger$\footnote{In this section $U\equiv U_{10-120}$ 
(or $U\equiv U_{126-120}$).} - any of the symmetry elements of $Y_{120}$, 
and $S_j^d$  which belongs to the symmetry of $Y_{10}^d$ (or $Y_{126}^d$):
\beq
\label{W120}
\left[W_{j \phi}^n\right]^p=\mathbb{I}\;\;\;,\;\;\;
W_{j\phi}^n = U g_\phi^n U^\dagger S_j^d. 
\eeq
Eq. (\ref{W120}) can be rewritten as
\beq
\label{aWexact}
{\rm Tr} \left[W_{j \phi}^n\right] = a_{p_n}(k_n,l_n), 
\eeq
where we will assume again that ${\rm Det} [W]=1$, 
so that the sum of the eigenvalues  equals  
\beq
\label{apkl}
a_{p_n}(k_n,l_n)=e^{2\pi i(k_n/p_n)}+e^{2\pi i(l_n/p_n)}+e^{-2\pi i(k_n/p_n+l_n/p_n)}. 
\eeq
Here all terms can differ from 1, so for a given $p_n$ the trace (\ref{aWexact}) is determined by two 
parameters $k_n$ and $l_n$. All inequivalent triples $(k/p,l/p,-(k+l)/p\;{\rm mod}\;1)$ for $k=1,\ldots,p-1$, 
$l=0,\ldots,p-1$) and $2\leq p\leq5$ with the corresponding $a_p(k,l)$ (see also \cite{Esmaili:2015pna}) 
are given in Table \ref{trojcki}.

%%%%%%%%%%%%%%%%%%%%%%%%%%%%%%%%%%%%%%%%%%%%%%%%%%%%%%%%%%%%%%%%%
\begin{center}
 \begin{tabular}{|c || c |} 
 \hline
$\left(\frac{k}{p},\frac{l}{p},-\frac{k+l}{p}\;{\rm mod}\;1\right)$ & $a_p(k,l)$  \\ [0.5ex] 
 \hline\hline
$\left(0,\frac{1}{2},\frac{1}{2}\right)$ & $-1$ \\ 
 \hline
$\left(0,\frac{1}{3},\frac{2}{3}\right)$ & $0$ \\ 
 \hline
$\left(\frac{1}{3},\frac{1}{3},\frac{1}{3}\right)$ & $-\frac{3}{2}+i\frac{3\sqrt{3}}{2}$\\ 
 \hline
$\left(\frac{2}{3},\frac{2}{3},\frac{2}{3}\right)$ & $-\frac{3}{2}-i\frac{3\sqrt{3}}{2}$ \\ 
 \hline
$\left(0,\frac{1}{4},\frac{3}{4}\right)$ & $1$ \\ 
 \hline
$\left(\frac{1}{4},\frac{1}{4},\frac{1}{2}\right)$ & $-1+i2$ \\ 
 \hline
$\left(\frac{1}{2},\frac{3}{4},\frac{3}{4}\right)$ & $-1-i2$ \\ 
 \hline
$\left(0,\frac{1}{5},\frac{4}{5}\right)$ & $\frac{1+\sqrt{5}}{2}$ \\ 
 \hline
$\left(\frac{1}{5},\frac{1}{5},\frac{3}{5}\right)$ & $-\frac{3-\sqrt{5}}{4}+i\sqrt{\frac{5(5-\sqrt{5})}{8}}$ \\ 
 \hline
$\left(\frac{1}{5},\frac{2}{5},\frac{2}{5}\right)$ & $-\frac{3+\sqrt{5}}{4}+i\sqrt{\frac{5(5+\sqrt{5})}{8}}$ \\ 
 \hline
$\left(0,\frac{2}{5},\frac{3}{5}\right)$ & $\frac{1-\sqrt{5}}{2}$ \\ 
 \hline
$\left(\frac{2}{5},\frac{4}{5},\frac{4}{5}\right)$ & $-\frac{3-\sqrt{5}}{4}-i\sqrt{\frac{5(5-\sqrt{5})}{8}}$ \\ 
 \hline
$\left(\frac{3}{5},\frac{3}{5},\frac{4}{5}\right)$ & $-\frac{3+\sqrt{5}}{4}+i\sqrt{\frac{5(5+\sqrt{5})}{8}}$ \\ 
 \hline
  \hline
\end{tabular}
\captionof{table}{\label{trojcki} Possible inequivalent values of $(k/p,l/p,-(k+l)/p\;{\rm mod}\;1)$ 
for $2\leq p\leq5$, $0\leq k\leq p-1$, $1\leq l\leq p-1$ with its corresponding $a_p(k,l)$.}
\end{center}
%%%%%%%%%%%%%%%%%%%%%%%%%%%%%%%%%%%%%%%%%%%%%%%%%%%%%%%%%%%%%%

Since now  ${\rm Tr} \left[W_j \right]$ is complex,  
Eq. (\ref{aWexact}) provides two relations on the mixing parameters for each $n$. 
For the real part we get 
\beq
\label{real}
{\rm Re}\left(a_{p_n}(k_n,l_n)\right)=
-1  + 2 \left|U_{j1}\right|^2 \left((1 - \cos{(2\pi n/p)}\right),  
\eeq
which depends on the absolute value $|U_{j1}|$ with the column index 1 and the latter 
is related to the form 
of $g(\vec{\phi})$ (\ref{eq:g120}) in which $g_{11}$ is isolated (decouples from the rest).
Changing place of this element to 22 or 33 will fix another column. 
Also interchanging $g$ and $S_j$ we can fix a row rather than  a column.

The imaginary part equals 
\beq
\label{imag}
{\rm Im}\left(a_{p_n}(k_n,l_n)\right)=
2\sin{(2\pi n/p)}\left(1-\left|U_{j1}\right|^2\right)\hat{e}\hat{\phi}
\eeq
with the unit vectors $\hat{e}$ and $\hat{\phi}$ defined as 
\bea
\label{e}
\hat{e}&=&
\frac{1}{1-\left|U_{j1}\right|^2}
\left[
2{\rm Re} \left(U_{j2}U_{j3}^*\right) , 
2{\rm  Im} \left(U_{j2}U_{j3}^*\right), 
\left|U_{j2}\right|^2-\left|U_{j3}\right|^2
\right],\\
\hat{\phi}& = &\vec\phi/\phi.
\eea
If $|U_{j1}|=1$ the r.h.s. of (\ref{imag}) vanishes.

There are thus $2\times(p-1)$ equations (\ref{real}) and (\ref{imag}) to solve, 
i.e. for all possible values of $n=1,\ldots,p-1$. This is possible only if $p_n,k_n,l_n$ depend 
on $n$. Essentially $|U_{j1}|$ can be found from Eq. (\ref{real}), while Eq. (\ref{imag}) provides 
a constraint on the angle $\hat e\hat \phi$. We will say more about possible solutions in 
section \ref{120}.

Notice that now the constraint on possible matrices $U$ found in 
\cite{Fonseca:2014koa,Fonseca:2015kya} is not valid, 
since in the case with $120_H$, the matrix element $|U|^2$ is not related to 
$k$ and $p$ only, as in (\ref{Uji}) or (\ref{Ujiminus}), but must satisfy more 
complicated equations (\ref{real})-(\ref{imag}).

Let us now give three examples involving the system with $120$.\\

As a first example consider the case of $p=4$. 
We thus have to find $(p-1)=3$ triples ($n=1,2,3$)
\beq
\label{T}
T_n\equiv\left(k_n,l_n,-(k_n+l_n)\;{\rm mod}\;p_n\right)/p_n
\eeq
which satisfy the $(p-1)=3$ equations (\ref{real}) and $p-1=3$ equations (\ref{imag}), 
allowing a solution for $|U_{j1}|$ and $\hat e \hat \phi$. 
An example of possible solution is given by 
\bea
\label{tripleexample}
n=1&\to&T_1=(1,1,2)/4\\
n=2&\to&T_2=(0,1,1)/2\\
n=3&\to&T_3=(2,3,3)/4
\eea
In fact it is easy to see explicitly that the ratios
\beq
\frac{{\rm Re}(a_{p_n}(k_n,l_n))+1}{1-\cos{(2\pi n/3)}}\;\;\;,\;\;\;
\frac{{\rm Im}(a_{p_n}(k_n,l_n))}{\sin{(2\pi n/3)}}
\eeq
are, for triples (\ref{tripleexample}), either undefined ($0/0$) or independent on $n$, giving 
$|U_{j1}|=0$ and $\hat e \hat \phi=1$. Other solutions of (\ref{aWexact}) will be given in 
section \ref{120}.\\

In the second example consider three Yukawa couplings
\beq
Y_{10}=U_{10}^*Y_{10}^dU_{10}^\dagger\;\;\;,\;\;\;
Y_{126}=U_{126}^*Y_{126}^dU_{126}^\dagger\;\;\;,\;\;\;
Y_{120}=U_{120}^*Y_{120}^dU_{120}^\dagger . 
\eeq
We assume that $G_f$ contains the following $p+1$ symmetry elements from these Yukawas:  
\beq
S_{10}= U_{10}S_i^dU_{10}^\dagger\;\;\;,\;\;\;
S_{126}= U_{126}S_i^dU_{126}^\dagger\;\;\;,\;\;\;
S_{120}^n=U_{120}g_\phi^n U_{120}^\dagger. 
\eeq
As for $g_\phi^n$ ($n=1,\ldots,p-1$) we should select a finite Abelian 
subgroup of the SU(2), $Z_p$, to embed into discrete $G_f$:
\beq
g_\phi^p=\mathbb{I}\to   \phi = \frac{\,2\pi}{p} . 
\eeq
Then the embedding of $S_{10}$, $S_{126}$ and $S_{120}^n$ into $G_f$ implies 
the symmetry group relations
\bea
\left(W_{ij}^{U_{10-126}}\right)^{p'}&  
\equiv & \left(U_{10-126}S_i^dU_{10-126}^\dagger S_j^d\right)^{p'} = \mathbb{I}, \\
\left(W_{j}^{U_{10-120}}\right)^{p''_n} & \equiv & \left(U_{10-120}g_\phi^n U_{10-120}^\dagger 
S_j^d\right)^{p''_n} = \mathbb{I}, \\
\label{126120}
\left(W_{i}^{U_{126-120}}\right)^{p'''_n} &  \equiv & \left(
U_{126-120} g_\phi^n  U_{126-120}^\dagger  S_i^d\right)^{p'''_n} = \mathbb{I}. 
\eea
They lead to the $4p-3$ real relations 
\bea
\label{equ126}
\left|\left(U_{10-126}\right)_{ji}\right|&=&\cos{\left(\pi\frac{k'}{p'}\right)}, \\
\label{equ120}
Tr\left[W_j^{U_{10-120}}\left(2\pi\frac{n}{p}\right)\right]&=&a_{p''_n}(k''_n,l''_n),\\
\label{equ20-26}
Tr\left[W_i^{U_{126-120}}\left(2\pi\frac{n}{p}\right)\right]&=&a_{p'''_n}(k'''_n,l'''_n). 
\eea
Eq. (\ref{equ126}) gives a bound on one element of $U_{10-126}$, 
eqs. (\ref{equ120}) -- on $U_{10-120}$,  
whereas eqs. (\ref{equ20-26}) on the product of the two: $U_{126-120}=U_{10-126}^\dagger U_{10-120}$. 
More precisely, from the real part of (\ref{equ120}) we obtain 
$|(U_{10-120})_{j1}|^2$, while the 
real part of (\ref{equ20-26}) gives 
\beq
\left|(U_{126-120})_{i1}\right|^2=\left|(U_{10-126}^*)_{ji}(U_{10-120})_{j1}+ 
\sum_{k\ne j}(U_{10-126}^*)_{ki}(U_{10-120})_{k1}\right|^2. 
\eeq
Imaginary parts  give constraints 
on $\hat \phi\,\hat e_{10-120}$ and $\hat \phi\,\hat e_{126-120}$ 
according to (\ref{imag}) with the definition (\ref{e}).
\\

In the third example  we consider a system with two Yukawas, 
e.g. $Y_{10}$ and $Y_{120}$. 
We can, similarly as in section \ref{unitarity}, see how unitarity restricts possible 
solutions when two (and thus due to group relations all three) among $S_j^d$ in (\ref{W120}) 
are residual symmetries. We thus have 
\bea
\label{W1}
Tr\left[W_1\left(2\pi\frac{n}{p}\right)\right]&=&a_{p'_n}(k'_n,l'_n),\\
\label{W2}
Tr\left[W_2\left(2\pi\frac{n}{p}\right)\right]&=&a_{p''_n}(k''_n,l''_n),\\
\label{W3}
Tr\left[W_3\left(2\pi\frac{n}{p}\right)\right]&=&a_{p'''_n}(k'''_n,l'''_n).
\eea
What we have to do is (restricting the solutions to $p,p',p'',p'''\leq5$) to find in Table
\ref{fromphi120} three solutions for the same $p$ with the sum 
\beq
\sum_{j=1}^3|(U_{10-120})_{j1}|^2=1.
\eeq
Up to permutations of elements we get
\bea
p=3&\to&\left|(U_{10-120})_{j1}\right|=\left(\frac{1}{\sqrt{3}},\frac{1}{\sqrt{3}},\frac{1}{\sqrt{3}}\right), 
\left(\sqrt{\frac{2}{3}},\frac{1}{\sqrt{3}},0\right), \left(\sqrt{\frac{3+\sqrt{5}}{6}}, 
\sqrt{\frac{3-\sqrt{5}}{6}}, 0\right)\non\\
p=4&\to&\left|(U_{10-120})_{j1}\right|=\left(\frac{1}{\sqrt{2}},\frac{1}{\sqrt{2}},0\right)\\
p=5&\to&\left|(U_{10-120})_{j1}\right|=\left(\sqrt{\frac{5+\sqrt{5}}{10}},\sqrt{\frac{5-\sqrt{5}}{10}},0\right).\non
\eea
We can ask if just unitarity is enough to get these solutions, repeating the arguments of section \ref{unitarity}. 
Summing the three equations (\ref{W1})-(\ref{W3}), we find the relation 
\beq
\label{sum1-3}
a_{p'_n}(k'_n,l'_n)+a_{p''_n}(k''_n,l''_n)+a_{p'''_n}(k'''_n,l'''_n)=-1-2\cos{\left(2\pi n/p\right)}.
\eeq
Although solving this equation (either by explicit numerical guess or using the techniques 
of \cite{ConwayJones}) is not problematic, one needs to combine $n=1,\ldots,p-1$ such 
solutions. In other words, satisfying the equation for the sum (\ref{sum1-3}) is necessary but, 
in general, not sufficient condition for solving the whole system (\ref{W1})-(\ref{W3}).

\section{\label{data}Confronting  relations with data}
%%%%%%%%%%%%%%%%%%%%%%%%%%%%%%%%%%%%%%%%%%%%%%%%%%%%%%%%%%%%%%%%%

The possible values of $\left|U_{ji}\right|$ found in sect. \ref{ressym}  
are of the form (\ref{Uji}). 
For $p\leq 5$ their values  are summarized in the Table \ref{pkU}. 
%%%%%%%%%%%%%%%%%%%%%%%%%%%%%%%%%%%%%%%%%%%%%%%%%%%%%%%%%%%%%%%%%
\begin{center}
 \begin{tabular}{||c | c | c | c||} 
 \hline
 $p$ & $k$ & $\left|\cos{\left(\pi k/p\right)}\right|$ & 
$\left|\sin{\left(\pi k/p\right)}\right|$ \\ [0.5ex] 
 \hline\hline
 2 & 1 & 0 & 1 \\ 
 \hline
 3 & 1 & 0.5 & 0.866 \\
 \hline
 4 & 1 & 0.707 & 0.707 \\
 \hline
 5 & 1 & 0.809 & 0.588 \\
 \hline
 5 & 2 & 0.309 & 0.951 \\ %[1ex] 
 \hline
\end{tabular}
\captionof{table}{\label{pkU} Possible values of $\left|U_{ji}\right|$ for $p\leq5$.}
\end{center}
%%%%%%%%%%%%%%%%%%%%%%%%%%%%%%%%%%%%%%%%%%%%%%%%%%%%%%%%%%%%%%%
Let us confront these values  with values extracted from the data.  
We start with (\ref{WYukawa}). The vacuum expectation 
values\footnote{Here we assume supersymmetry;  
in the non-supersymmetric case, the Higgs 10-plet 
and 120-plet are in principle real. In this case 
$v_{10,120}^d=\left(v_{10,120}^u\right)^*, w_{120}^d=\left(w_{120}^u\right)^*$.} (VEVs) 
$v_{10,120}^{u,d}$, $w_{126,120}^{u,d}$ of the $10_H, \overline{126}_H, 120_H$ Higgses 
break SU(2)$_L\times$U(1)$_Y\to$U(1)$_{em}$ 
and generate the mass matrices for up quark, 
down quark, charged leptons, and Dirac neutrinos correspondingly: 
\bea
M_U&=&v_{10}^uY_{10}+w_{126}^uY_{126}+\left(v_{120}^u+w_{120}^u\right)Y_{120},
\nonumber\\
M_D&=&v_{10}^dY_{10}+w_{126}^dY_{126}+\left(v_{120}^d+w_{120}^d\right)Y_{120},
\nonumber\\
M_E&=&v_{10}^dY_{10}-3w_{126}^dY_{126}+\left(v_{120}^d-3w_{120}^d\right)Y_{120},
\nonumber\\
M_{\nu_D}&=&v_{10}^uY_{10}-3w_{126}^uY_{126}+\left(v_{120}^u-3w_{120}^u\right)Y_{120}
\label{mass-yuk}
\eea
The non-zero neutrino mass comes from both type I and II contributions:
\beq
M_N=-M_{\nu_D}^TM_{\nu_R}^{-1}M_{\nu_D}+M_{\nu_L}, 
\eeq
where the left-handed, $M_{\nu_L}$, and right-handed, $M_{\nu_R}$, Majorana mass matrices 
are generated by non-vanishing 
(in the Pati-Salam decomposition) SU(2)$_R$ triplet VEV $v_R$ 
and SU(2)$_L$ triplet VEV $v_L$:
\beq
M_{\nu_L}=v_LY_{126}\;,\;\;\;
M_{\nu_R}=v_RY_{126}. 
\label{leftright}
\eeq

Relations (\ref{mass-yuk}) - (\ref{leftright}) and the experimental 
values of the SM fermion masses and mixing
allow to reconstruct (with some additional assumptions) 
the values of the Yukawa matrices $Y_{10}$  and $Y_{126}$.  
Then diagonalizing these matrices as in (\ref{uadiagon})
we can get the relative matrices, e.g. $U_{10-126} = U_{10}^{\dagger} U_{126}$. 
The procedure of reconstruction of $Y_{a}$ from the data is by far not unique 
and a number of assumptions and further restrictions are needed to get $Y_a$. 
Here we will  describe few cases from the literature, where the 
unitary matrices $U_a$ are explicitly given. For other 
fits see for example \cite{fitminSO10GUT}.

\subsection{\label{checksym}The case of $Y_{10}+Y_{126}$}
%%%%%%%%%%%%%%%%%%%%%%%%%%%%%%%%%%%%%%%%%%%%%%%%%%%%%%%%%%%%%%%%%

Consider first check whether equality (\ref{Uji}) 
is satisfied for one or more elements of the 
reconstructed  relative matrix $U_{10-126}$.  
A fit of the Yukawas has been done, for example,  in \cite{Dueck:2013gca}, 
where the SUSY scale was assumed to be low. 
Let us start with the Yukawas displayed in eq. (18) of \cite{Dueck:2013gca}. The 
corresponding matrix $U$ (only absolute values of its elements are important) 
can be found easily:
\beq
\left|U_{10-126}\right|=
\left(
\begin{array}{ccc}
 0.919 & 0.392 & 0.037 \\
 0.362 & 0.812 & 0.458 \\
 0.156 & 0.432 & 0.888
\end{array}
\right). 
\label{mmm1}
\eeq
One should also take into account possible uncertainties in 
the determination  of elements of (\ref{mmm1}), which we estimate as $10-20\%$.  
The element $\left| (U_{10-126})_{22} \right|$ is numerically close to 
$\left|\cos{\left(\pi/5 \right)} \right|$. Furthermore,  
$|(U_{10-126})_{23}|=0.46\approx0.5=\cos{\left(\pi/3\right)}$. The third element in the 
same row is $|(U_{10-126})_{21}|=0.36\approx0.31=\cos{\left(2\pi /5\right)}$. 
This is one of the cases in which a full row of the relative matrix 
is determined  by a residual symmetry, namely by  the solution in (\ref{ccc2}). 
One can interpret this as an experimental evidence for the existence of $G_f$. 

The second example comes from the Yukawa couplings shown in eq. (22) 
of \cite{Dueck:2013gca}. They lead to  the relative mixing matrix 
\beq
\left|U_{10-126}\right|=
\left(
\begin{array}{ccc}
 0.958 & 0.285 & 0.033 \\
 0.262 & 0.917 & 0.301 \\
 0.116 & 0.280 & 0.953
\end{array}
\right). 
\eeq
The matrix element $\left|(U_{10-126})_{23}\right|$ is numerically close to 
$\left|\cos{\left(2\pi/5\right)}\right|$, however the other 
elements in the same row or column are not close to any   
value determined by symmetry. 
With large probability this can be just accidental coincidence. 

\subsection{\label{120}Relative mixing between $Y_{10}$ and  $Y_{120}$}
%%%%%%%%%%%%%%%%%%%%%%%%%%%%%%%%%%%%%%%%%%%%%%%%%%%%%%%%%%%%%%%%%

Let us check if the elements of  the relative matrix 
$U_{10-120} = U_{10}^\dagger U_{120}$
are  in agreement with data   
for some choice of $j$, $p$ and 
\beq
T_n\equiv\left(\frac{k_n}{p_n},\frac{l_n}{p_n},-\frac{k_n+l_n}{p_n}\;{\rm mod}\;1\right)\;\;\;,\;\;\;n=1,\ldots,p-1 .
\eeq
Taking different values for $p$ and $T_n$, 
we predict  $|(U_{10-120})_{j1}|$. All possible values of 
$|(U_{10-120})_{j1}|$  and corresponding $\hat e\hat\phi$, 
for $p,p_n = 2,3,4,5$ are shown in Table \ref{fromphi120}.
They  are solutions of eqs. (\ref{real})-(\ref{imag}).  

%%%%%%%%%%%%%%%%%%%%%%%%%%%%%%%%%%%%%%%%%%%%%%%%%%%%%%%%%%%%%%%%%
\begin{center}
\begin{tabular}{| c | c | c | c | c || c | c |} 
 \hline
$p$ & $T_1$ & $T_2$ & $T_3$ & $T_4$ & $|(U_{10-120})_{j1}|$ & $\hat e\hat \phi$ \\ [0.5ex] 
 \hline\hline
\multirow{5}{0.5em}{$3$} & $\left(0,\frac{1}{2},\frac{1}{2}\right)$ & $\left(0,\frac{1}{2},\frac{1}{2}\right)$ & - & - &
$0$ & $0$ \\
& $\left(0,\frac{1}{3},\frac{2}{3}\right)$ & $\left(0,\frac{1}{3},\frac{2}{3}\right)$ & - & - & 
$\sqrt{\frac{1}{3}}=0.577$ & $0$ \\
& $\left(0,\frac{1}{4},\frac{3}{4}\right)$ & $\left(0,\frac{1}{4},\frac{3}{4}\right)$ & - & - & 
$\sqrt{\frac{2}{3}}=0.816$ & $0$ \\
& $\left(0,\frac{1}{5},\frac{4}{5}\right)$ & $\left(0,\frac{1}{5},\frac{4}{5}\right)$ & - & - & 
$\sqrt{\frac{3+\sqrt{5}}{6}}=0.934$ & $0$ \\
& $\left(0,\frac{2}{5},\frac{3}{5}\right)$ & $\left(0,\frac{2}{5},\frac{3}{5}\right)$ & - & - & 
$\sqrt{\frac{3-\sqrt{5}}{6}}=0.357$ & $0$ \\
  \hline
\multirow{4}{0.5em}{$4$} & $\left(0,\frac{1}{2},\frac{1}{2}\right)$ & $\left(0,\frac{1}{2},\frac{1}{2}\right)$ & 
$\left(0,\frac{1}{2},\frac{1}{2}\right)$ & - & $0$ & $0$ \\
& $\left(0,\frac{1}{3},\frac{2}{3}\right)$ & $\left(0,\frac{1}{4},\frac{3}{4}\right)$ & 
$\left(0,\frac{1}{3},\frac{2}{3}\right)$ & - & $\sqrt{\frac{1}{2}}=0.707$ & $0$ \\
& $\left(\frac{1}{4},\frac{1}{4},\frac{1}{2}\right)$ & $\left(0,\frac{1}{2},\frac{1}{2}\right)$ & 
$\left(\frac{1}{2},\frac{3}{4},\frac{3}{4}\right)$ & - & $0$ & $1$ \\
& $\left(\frac{1}{2},\frac{3}{4},\frac{3}{4}\right)$ & $\left(0,\frac{1}{2},\frac{1}{2}\right)$ & 
$\left(\frac{1}{4},\frac{1}{4},\frac{1}{2}\right)$ & - & $0$ & $-1$ \\
  \hline
\multirow{4}{0.5em}{$5$} & $\left(0,\frac{1}{2},\frac{1}{2}\right)$ & $\left(0,\frac{1}{2},\frac{1}{2}\right)$ & 
$\left(0,\frac{1}{2},\frac{1}{2}\right)$ & $\left(0,\frac{1}{2},\frac{1}{2}\right)$ & $0$ & $0$ \\
& $\left(0,\frac{1}{3},\frac{2}{3}\right)$ & $\left(0,\frac{1}{5},\frac{4}{5}\right)$ & 
$\left(0,\frac{1}{5},\frac{4}{5}\right)$ & $\left(0,\frac{1}{3},\frac{2}{3}\right)$ & $\sqrt{\frac{5+\sqrt{5}}{10}}=0.851$ & $0$ \\
& $\left(0,\frac{2}{5},\frac{3}{5}\right)$ & $\left(0,\frac{1}{3},\frac{2}{3}\right)$ & 
$\left(0,\frac{1}{3},\frac{2}{3}\right)$ & $\left(0,\frac{2}{5},\frac{3}{5}\right)$ & $\sqrt{\frac{5-\sqrt{5}}{10}}=0.526$ & $0$ \\
  \hline
\end{tabular}
\captionof{table}
{\label{fromphi120} Predictions for $|(U_{10-120})_{j1}|$ for   
$p,p_n= 2, 3, 4, 5$. 
The outputs, solutions of (\ref{real})-(\ref{imag}), 
are $|(U_{10-120})_{j1}|$ and $\hat e\hat \phi$. }
\end{center}
%%%%%%%%%%%%%%%%%%%%%%%%%%%%%%%%%%%%%%%%%%%%%%%%%%%%%%%%%%%%%%%%%
We reconstruct  $U_{10-120}$  from the Table 2  p. 39 of \cite{Aulakh:2013lxa}: 
\beq
\label{absU120}
|U_{10-120}|=
\left(
\begin{array}{ccc}
 0.951 & 0.310 & 0 \\
 0.306 & 0.939 & 0.158 \\
 0.049 & 0.150 & 0.987 \\
\end{array}
\right). 
\eeq
Confronting  the first column in 
this matrix with predictions of the Table \ref{fromphi120} 
we find 
that $|(U_{10-120})_{11}| = 0.951$  is close to one of the five solutions 
for $p=3$: $\sqrt{(3+\sqrt{5})/6} = 0.934 $. 

Other data fits give substantially different matrices  $U_{10-120}$.   
The following  values for the elements of the 
first columns of $U_{10-120}$ have been found \footnote{We 
thank Charanjit Khosa for these data.}
\beq
|(U_{10 -120})_{j1}|=\left(
\begin{array}{c}
 0.865 \\
 0.490 \\
 0.113 \\
\end{array}
\right),\;
\left(
\begin{array}{c}
 0.828 \\
 0.540 \\
 0.150 \\
\end{array}
\right),\;
\left(
\begin{array}{c}
 0.928 \\
 0.354 \\
 0.117 \\
\end{array}
\right),\;
\left(
\begin{array}{c}
 0.640 \\
 0.753 \\
 0.155 \\
\end{array}
\right). 
\eeq
Again, coincidences with predictions of the Table \ref{fromphi120} can be found. 

\subsection{RG invariance of the residual symmetry}
%%%%%%%%%%%%%%%%%%%%%%%%%%%%%%%%%%%%%%%%%%%%%%%%%%%%%%%%%%%%%%%%%

Since we consider here the symmetry at the SO(10) level, 
the relative mixing matrix $U_{b-a}$, 
determined by the residual symmetries, should be considered at GUT or even higher mass 
scales. One would expect that 
renormalization group equation running change the value of this unitary matrix. 
This, indeed, happens in most of the cases,  for example when 
residual symmetries are applied to 
quarks or leptons in the standard model: 
the CKM or PMNS matrices run, so that the validity of the residual symmetry approach 
is bounded to an a-priori unknown scale. 

In any supersymmetric SO(10)  a residual 
symmetry inposed at the GUT scale will remain such also at any scale above it. 
Indeed, due to supersymmetry the renormalization 
is coming only through wave-functions. This means that up to 
wave-function renormalization of the $10_H$ and $\overline{126}_H$ the Yukawa matrices 
$Y_{10}$ and $Y_{126}$ above the GUT scale renormalize in the same way:
\beq
\left(Y_{10}^{ren}\right)_{ij}=\left(Z_{16}\right)_{ii'}\left(Z_{16}\right)_{jj'}Z_{10}
\left(Y_{10}\right)_{i'j'}, \\
\left(Y_{126}^{ren}\right)_{ij}=\left(Z_{16}\right)_{ii'}\left(Z_{16}\right)_{jj'}Z_{126}
\left(Y_{126}\right)_{i'j'}. 
\eeq
The different renormalization of (different) Higgses $H_a$
gives just an overall factors, and as such appears as a common multiplication 
the corresponding Yukawa matrices $Y_a$, without 
change of the relative mixing matrix $U_{b-a}$. 
This  is different from other cases, where a residual symmetry 
is valid at a single scale only. Here if the symmetry 
exists at the SO(10) GUT scale, it is 
present also at any scale above it, thanks to the combined effect of supersymmetry and SO(10).

%%%%%%%%%%%%%%%%%%%%%%%%%%%%%%%%%%%%%%%%%%%%%%%%%%%%%%%%%%%%%%%%%
\section{\label{singlets}SO(10) model with hidden sector}
%%%%%%%%%%%%%%%%%%%%%%%%%%%%%%%%%%%%%%%%%%%%%%%%%%%%%%%%%%%%%%%%%

Another class of SO(10) models includes the SO(10) fermionic singlets $S$ which 
mix with the usual neutrinos via the Yukawa couplings with $16_H$ 
(see \cite{Chu:2016lkb} and references therein). 
This avoids the introduction of high dimensional Higgs representations 
$126_H$  and $120_H$ to generate fermion masses. 
Neutrino masses are generated via the double seesaw \cite{Mohapatra:1986bd}
and this allows to disentangle generation of the quark mixing and lepton mixing, 
and therefore naturally explain their different patterns.  
The Lagrangian of the Yukawa sector is given by   
\beq
\label{WYukawa2}
{\cal L}_{Yukawa} = 16_F^T 10_H^q Y_{10}^q 16_F + 
  16_F^T Y_{16} 16_H  S + S^T  Y_{1} 1_H S + ..., 
\eeq
where subscripts  $q = u,~ d$ refer to different 
Higgs 10-plets. The  matrices  of Yukawa couplings, $Y_{10}$, $Y_{16}$ and $Y_{1}$
correspond to  Higgses in $10_H$, ${16}_H$ and $1_H$. 
If not suppressed by symmetry, the singlets may have also the bare mass terms. 
Additional interactions should be added to (\ref{WYukawa2}) to explain the 
difference of mass hierarchies of quarks and charged leptons. 
Two $10$ plets of Higgses can be introduced to generate  different 
mass scales of the upper and down quarks. (Equality $Y_D=Y_E$
can  be broken by high order operators.) In these models the couplings of  $16_F$ 
with singlets (\ref{WYukawa2}) are responsible for the 
difference of mixing of quarks and leptons and for the smallness of neutrino masses. 

The Lagrangian (\ref{WYukawa2}) contains three fermionic operators of 
different SO(10)  structure  $16_F 16_F$, $16_F 1_F$ and $1_F 1_F$   
in contrast to (\ref{WYukawa}),  where 
all the terms have  the same $16_F 16_F$ structure. This also can be an origin 
of different symmetries of $Y_a$ on the top of difference of Higgs representations. 

The terms in (\ref{WYukawa2}) have different intrinsic symmetries: 

1. The first one has the Klein group symmetry $G_{10} = Z_2 \times Z_2$,  as  
the terms in (\ref{WYukawa}).  

2. The last term is also symmetric and  has $G_{1} = Z_2 \times Z_2$ symmetry. 

3. The second, ``portal'' term obeys  a much wider intrinsic symmetry: 
$U(1) \times U(1) \times U(1)$. In the diagonal  basis it is related to independent 
continuous rotation of the three diagonalized states. This term can be considered as the Dirac 
term of charged leptons in previous studies of residual symmetries. 
To further proceed with the discrete symmetry approach we can select the discrete 
subgroup of the continuous symmetry, e.g. $G_{16} = Z_m \times  Z_n \times  Z_l$, 
or (to match with previous considerations in literature) even single subgroup $G_{16} = Z_n$, 
under which different components have different charges 
$k = 0, 1, ... n - 1$. So, the symmetry transformation, $T$,  in the diagonal basis  
$16_F' = T ~16_F$, $S' = T^\dagger ~S$ becomes:   
\beq
T  =
\bem
e^{i 2 \pi \frac{k_1}{n}}      & 0  &  0 \cr
0               & e^{i 2 \pi \frac{k_2}{n}}  &  0 \cr
0      &    0        &  e^{i 2 \pi \frac{k_3}{n}} 
\eem
\eeq
with $k_1 + k_2 + k_3 = 0\;{\rm mod}\;n$ to keep ${\rm Det} (T) = 1$. 

There are many possible embeddings of  the residual symmetries $G_{10}$,  $G_{16}$,  $G_{1}$
which will lead to restriction on the relative mixing matrices between 
$Y_{10}$,  $Y_{16}$,  $Y_{1}$. These  matrices will determine 
eventually the lepton mixing (and more precisely its difference from the quark mixing). 
Recall that the difference may have special form like TBM or BM-type. 

According to the double seesaw  \cite{Mohapatra:1986bd}  the light neutrino mass matrix equals 
\beq
m_\nu \propto Y_{10}^u~ Y_{16}^{T -1}~ Y_1~ Y_{16}^{-1}~ Y_{10}^{u T}. 
\eeq
In terms of the diagonal matrices and relative rotations it can be rewritten as 
\beq
m_\nu \propto Y_{10}^d~ U_{10-16}~ Y_{16}^{d -1}~ U_{16-1}~ 
Y_1^d ~U_{16-1}^T  ~Y_{16}^{-1}~ U_{10-16}^T  ~ Y_{10}^T. 
\eeq
Then the embedding of $G_{10}$ and   $G_{16}$ (or their subgroups) into a 
unique flavor group $G_f$ determines (restricts) 
the relative matrix $U_{10-16}$. Embedding of $G_{16}$ and $G_{1}$  
into $G_f'$ determines $U_{16-1}$. Further embedding of all 
residual symmetries will restrict both $U_{10-16}$  and  $U_{16-1}$. 

Let us mention one possibility. Selecting the parameters of embedding one can, e.g. 
obtain $U_{10-16} = \mathbb{I}$ and 
$U_{16-1} = U_{TBM}$. Then imposing $Y_{10}^d Y_{16}^{d -1} = \mathbb{I}$ 
(which would require some additional symmetries \cite{Hagedorn:2008bc} \cite{Chu:2016lkb})
one finds
\beq
m_\nu =  U_{16-1} ~Y_1^d ~U_{16-1}^T = U_{TBM}~ Y_1^d ~ U_{TBM}^T, 
\eeq
that is,  the TBM mixing of neutrinos.  
Detailed study of these possibilities is beyond the scope of this paper. 

\section{\label{intrinsicmixing}Intrinsic symmetries and relative mixing matrix}
%%%%%%%%%%%%%%%%%%%%%%%%%%%%%%%%%%%%%%%%%%%%%%%%%%%%%%%%%%%%%

Let us further clarify the conceptual issues related to
the intrinsic and residual symmetries.

Intrinsic symmetries are the symmetries left after breaking of
a bigger flavor symmetry. These symmetries exist  before and after
$G_f$ breaking. By itself these symmetries do not carry 
any new  information about the flavor apart from that of 
symmetricity of antisymmetricity of the Yukawa matrices.
So,  by itself the  intrinsic symmetries do not restrict
the flavor structure.

These symmetries do not depend on the model parameters or on symmetry breaking.
Recall that depending on the basis the form of symmetry transformation
is different. So, changing the basis leads to the change of the form.

In a given basis  symmetry transformations for different $Y_{a}$
can have different form, and it is this form of the transformation
that encodes the flavor information. In other words,
not the symmetry elements (generators) themselves, 
but their form in a given (and the same for all couplings) 
basis that encodes (restricts) the flavor structure. Changing basis
for all couplings simultaneously does not change physics.

Breaking of the flavor symmetry fixes the form of the
intrinsic symmetry transformations. In other words, $G_f$ breaking
can not  break the intrinsic symmetries but
determine the form of symmetry transformations
in a fixed (for all the couplings) basis.

In a sense, the intrinsic symmetries can be considered as a tool
to introduce the flavor symmetries and study their consequences. 
Indeed, in the usual consideration symmetry determines the 
form of the Yukawa matrices in a certain basis. Changing the basis
leads to a change of the form of $Y_a$, but it does not change
the relative mixing matrix between different $Y_a$, 
which has a physical meaning. 
On the other hand the form of $Y_a$ determines the form of
symmetry transformations. Therefore studying the 
form of transformations we obtain consequences of symmetry.

Let us show that the matrix which diagonalizes $Y_a$
determines the form of symmetry transformation.
For definiteness we consider two symmetric  matrices $Y_a$ and $Y_b$,
and take the basis where $Y_b$ is diagonal.
The diagonalization of $Y_a$ in this basis is given by
rotation $U$:
\beq
Y_a = U^* Y_a^d U^\dagger.
\label{uadiagon}
\eeq
(Recall that here $U$ is the relative
mixing matrix $U_{b-a}$ and we omit subscript for brevity). 
Let us show that $U$ determines the form of the intrinsic symmetry transformation as
\beq
S = U S^d U^\dagger,
\label{sbasis}
\eeq
where $S^d$ is the intrinsic symmetry transformation in
the basis where $Y_a$ is diagonal: 
\beq
S^{d} Y_a^d S^d = Y_a^d.
\label{udiagg}
\eeq
Using (\ref{sbasis}) and (\ref{uadiagon}) we have
\beq
S^T Y_a S = U^* S^d U^T U^* Y^d_a U^\dagger U S^d U^\dagger
= U^* S^d Y^d_a S^d U^\dagger = U^* Y^d_a U^\dagger = Y_a, 
\label{yayaya}
\eeq
where in the second equality we used the invariance 
(\ref{udiagg}). 
According to  (\ref{yayaya}) $S$ defined in (\ref{sbasis})
is indeed the symmetry transformation of $Y_a$.\\ 

Let us comment on intrinsic and residual symmetries.
Not all intrinsic symmetries can be taken as residual symmetry
which originate from a given flavor symmetries.
On the other hand,  residual symmetries
can be bigger than just intrinsic symmetries,
i.e. include elements which are  not intrinsic. 
The variety of residual transformations does not
coincide with the variety of intrinsic symmetry transformations.\\ 

We can consider another class of symmetries under which  also the Higgs 
bosons are charged. The symmetries are broken by these Higgs 
VEVs. In the case of 
a single Higgs multiplet of a given dimension, this does not produce flavor structure. 

Let us comment on possible realization and  implications 
of the residual symmetry approach. We can assume that three 
$16_F$ form a triplet of the covering group  $G_f$  ($A_4$ 
can be taken as an example). If we assume that Higgs multiplets 
$H_a$, $a = 10, \overline{126}, 120, 16$,  
are singlets of $G_f$, then the product  $16_F^T Y_a 16_F$ should originate 
from $G_f$ symmetric interactions. Apart from trivial case 
of $Y_a \propto I$ (implied that $16_F^T 16_F$ is invariant under $G_f$), 
$Y_a$ should be the effective coupling that appears after spontaneous 
symmetry breaking, so it is the function of the flavon fields $\phi$, 
$\xi$,  which transform non-trivially under $G_f$: 
$Y_a = Y_a (\phi,\xi)$. In the $A_4$ example we may have, e.g., that 
\beq     
Y_{10} = h_{10}~y (\vec{\phi}), ~~~~ Y_{126} = h_{126}~ y (\xi), 
\eeq
where $\vec{\phi} = (\phi, \phi', \phi'')$ are flavons transforming as $1, 1', 1''$ 
representations of $A_4$ and $\xi$ transforms as a triplet of  $A_4$. 
The effective Yukawa couplings are generated when the flavons get VEV's. 
Then $Y_{10}$ will be diagonal, whereas $Y_{126}$ - off-diagonal. 

To associate $10_H$ with certain flavons we need to introduce another symmetry 
in such a way that only $\phi 10_H$ and $\xi 126_H$ are invariant.  
For instance,  we can introduce a $Z_4$ symmetry under which 
$\phi$,  $10_H$,  $\xi$,  $126_H$ transform with  $-1,~ -1,~ i, ~-i$, respectively.  

\section{\label{outlook}Summary and Conclusion}
%%%%%%%%%%%%%%%%ssss5%%%%%%%%%%%%%%%%%%%%%%%%%%%%%%%%%%%%%%%%%%%%%%

We have explored an interplay of the vertical (gauge) symmetry 
and flavor symmetries in obtaining the fermion masses and mixing. 
In SO(10) the GUT Yukawa couplings have intrinsic flavor symmetries 
related to the SO(10) gauge structure. These symmetries are always  
present independently of the specific parameters of the model 
(couplings or masses).  
Different terms of the Yukawa Lagrangian  have different intrinsic symmetries. 
Due to SO(10) the matrices of Yukawa couplings of $16_F$ with  the $10_H$ and $126_H$ 
are  symmetric and therefore have ``built-in'' $G_{10} =  Z_2\times Z_2$ 
and  $G_{126} =  Z_2\times Z_2$ symmetries.  
We find that the matrix of Yukawa couplings of $120_H$, being antisymmetric, 
has $G_{120} = SU(2)$ symmetry and some elements of the discrete subgroup 
of $SU(2)$ can be used for 
further constructions. If also SO(10) fermionic singlets $S$ exist,  
their self couplings are symmetric and therefore $G_{1} =  Z_2\times Z_2$. 
The couplings of $S$ with $16_F$ have symmetries of the Dirac type 
$G_{16} = U(1)^3$,  and the interesting subgroup is $G_{16} = Z_n$. \\ 

We assume that (part of) the intrinsic (built-in) symmetries 
are residual  symmetries which are left out from the breaking of a bigger flavor symmetry group 
$G_f$ \cite{Lam:2014kga}. So $G_f$  is the covering group 
of the selected residual symmetry groups.  
This is an extension of the residual symmetry approach used in the past to explain lepton  
mixing. The main difference is that in the latter case the 
mass terms with different residual symmetries involve different fermionic 
fields: neutrino and charged leptons. Here the Yukawa interactions 
with different symmetries involve the same $16_F$ 
(but different Higgs representations). 
Higgses are uncharged with respect to 
the residual symmetries but should encode somehow information about the Yukawa 
couplings. In the presence of the fermionic singlets, also the fermionic 
operators can encode this information. \\

We show that the embedding of the residual symmetries leads to  
determination of the  elements of the relative mixing matrix $U_{a-b}$
which connects the diagonal bases of the Yukawa matrices $Y_a$ and $Y_b$. 
In our analysis we use the symmetry group condition 
which allows  to determine the elements of the relative matrix immediately 
without the explicit construction of the Yukawa matrices  
and their diagonalization. We show the equivalence of our approach and the one in  
\cite{Lam:2014kga} in few explicit examples.\\

In the case of the minimal SO(10) with one $10_H$ and one $126_H$   
the total intrinsic symmetry is 
$G_{10}\times G_{126} = (Z_2\times Z_2)_{10} \times (Z_2\times Z_2)_{126}$. 
In this case the covering group is the Coxeter group. 
If one $Z_2$ element of  $G_{10}$ and 
one element of $G_{126}$ are taken,  
so that  the residual symmetry is $Z_2 \times Z_2$, 
only one element of the relative mixing matrix  $U_{10-126}$   
is determined.  The value of the element is given by the integers $p$, $k$ of the embedding  
and therefore  has a discrete ambiguity. 

If one $Z_2$ element of $G_{10}$ (or $G_{126}$ )  
and both  elements  of $G_{126}$ (or $G_{10}$) are taken 
as the residual symmetries, then two elements in a row (column) are 
determined. Furthermore, as a consequence of unitarity, the whole row (column) 
is determined. We show that the unitarity condition emerges from the group properties. Unitarity is not 
automatic and it  imposes additional conditions on the parameters of the embedding, 
and therefore on possible values of the matrix elements. 

If all $Z_2$ elements of $G_{10}$ and $G_{126}$  are taken  
as the residual symmetries, then 4 elements of $U$, and consequently,  the whole  
matrix $U$ is determined. \\

Using elements of $G_{120}$ opens up different possibilities. 
Taking the Abelian $Z_p$ subgroup of $SU(2)$ the covering group is not a 
Coxeter group anymore for $p>2$, and so not covered by \cite{Lam:2014kga}. 
Even if we start with one single element of $g\in Z_p$ ($g^p=\mathbb{I}$) being a residual symmetry, 
so must be $g^2,\ldots,g^{p-1}$. This follows simply from the definition of 
a group, it is not our choice or assumption. So each of the elements $g^n$, $n=1,\ldots,p-1$, 
must satisfy a group condition if also a $Z_2$ element of $G_{10}$ is a residual symmetry. 
In the case of residual symmetry 
with one $Z_2$ element of $G_{10}$ and the $p-1$ elements of 
$Z_p$ a total of $2\times(p-1)$ 
real equations for one element $|(U_{10-120})_{j1}|$ and one angle 
$\hat e\hat \phi$ (plus various integers) must be satisfied. Solutions can exist only because 
each complex equation can have a different choice of $p_n,k_n,l_n$. 

Using unitarity $3\times2\times(p-1)$ relations on elements of $U_{10-120}$ (plus some integers) 
appear if the whole $G_{10}$ and $p-1$ elements from $G_{120} = Z_p$ are taken as residual symmetries. 

If one $Z_2$ from $G_{10}$, another $Z_2$  from $G_{126}$ and 
$p-1$ elements $Z_p$  from $G_{120}$ are identified as  the residual symmetries, we obtain 
relations between the  elements of both $U_{10-126}$  and $U_{10-120}$. \\

We confronted the obtained values of  elements of the relative 
mixing matrices with available results of data fits. 
We find that in the case of $G_{10}$ and $G_{126}$ embedding  
the predictions for one and two elements are 
compatible with  some  fits. Also for $G_{10}$ and $G_{120}$ embeddings 
some predictions for elements of $U_{10-120}$ exactly or approximately coincide with data. 
These values as well as residual symmetries in general are renormalization group independent 
in supersymmetric SO(10). 

The fits to data are not unique and typically 
several local minima with low enough $\chi^2$  exist. This is one of 
the reasons why we cannot conclude yet that SO(10) data point toward 
residual symmetries, and more work should be done. The other reason 
is the unavoidable possibility that a coincidence between
data and the theoretical expectation could be simply accidental. \\

\section*{Acknowledgments}
The authors would like to express a special thanks to the Mainz Institute for Theoretical Physics 
(MITP) for its hospitality and support. BB would like to thank Pritibhajan Byakti, Claudia Hagedorn and 
Palash Pal for discussion, Charan Aulakh and Charanjit Khosa for discussion, correspondence and for 
sharing unpublished data. The work of BB has been supported by the Slovenian Research Agency.

\appendix 

\section{\label{Lam}Comparison with the approach in \cite{Lam:2014kga}}
%%%%%%%%%%%%%%%%%%%%%%%%%%%%%%%%%%%%%%%%%%%%%%%%%%%%%%%%%%%

The invariance of the symmetric Yukawa matrix $Y$ is expressed as 
\beq
S^T Y S = Y. 
\eeq
The intrinsic symmetry can be easily realized in the  basis
where the Yukawa matrix  $Y$ is diagonal. 
A diagonal matrix 
$Y^d =  diag (y_1, y_2, y_3)$
with arbitrary (non-degenerate) elements $y_i$ 
is invariant with respect to transformations 
\beq
S_i^d =  diag [ (-1)^n,  (-1)^k,  (-1)^l ], ~~~~ n, k, l, = 0, 1.    
\label{eqtransall}
\eeq
For a symmetric matrix the invariance is defined as 
\beq
S_i^d Y^d  S_i^d = Y^d. 
\eeq
The elements, being  reflections, satisfy $(S_i^d)^2 = \mathbb{I}$.  
There are $2^3 = 8$ different transformations in (\ref{eqtransall}), 
including the identity matrix. So, the maximal intrinsic symmetry group is 
$Z_2^3$, since transformations with  $s_j = - s_i$ having  opposite signs  
of determinants,   do not produce additional restrictions on $m$.  
If we take elements with ${\rm Det} (S_i) = 1$,  only 4 elements 
are left which correspond to the $Z_2 \times Z_2$ group. 

In general, different Yukawa matrices can not be diagonalized simultaneously.  
Therefore, in a given basis, their symmetry elements  
can be obtained performing the unitary transformation:  
\beq
S_i = U_i S_i^d U_i^\dagger,  
\label{arbbasis}
\eeq
where $U_i$ connects  a given basis with the diagonal basis for $S_i$. 
Using  $(S_i)^2 = \mathbb{I}$ it is easy to show that  
for two different elements $(S_i S_j)^n = (S_j S_i)^n$ 
with  $n \geq 2$.  The group formed by the reflection  elements $S_i$ is called 
the Coxeter group\footnote{A Coxeter group with two generators is a 
von Dyck group $D(2, 2, p)$.}. 

In \cite{Lam:2014kga} it is suggested that 
different terms of the SO(10) Yukawa Lagrangian, and consequently different 
mass matrices generated by these terms,   are invariant 
under different elements $S_i$. Furthermore     
$S_i$ are identified with the residual symmetry  left over from  the breaking of the Coxeter group. 
Invariance of the Yukawa matrices leads to restriction of their elements.\\ 

Let us show that the approach in this paper is equivalent to that in  \cite{Lam:2014kga}. 
Consider two Yukawa matrices (or ``fundamental" mass  matrices as in  \cite{Lam:2014kga}) 
$Y_a$ and  $Y_b$ invariant with respect to $S_a$ and $S_b$. Then the elements  $S_a$ and $S_b$
being  residual symmetry elements satisfy the relation $(S_a S_b)^p = \mathbb{I}$. Expressing $S_a$ and $S_b$ 
in terms  of diagonal elements (\ref{arbbasis}) we obtain  
$(U_{a-b} S_a^d U^\dagger_{a-b} S_b^d)^p =  \mathbb{I}$,  where $U_{a-b} \equiv U_b^\dagger U_a$
This coincides with the symmetry group condition (\ref{Wijpje1}). 
$U_{a-b}$ connects two basis in which $Y_a$, $Y_b$ are diagonal,  
that is, the relative mixing matrix. This matrix does not depend on the basis and has 
a physical meaning. \\

In our approach we use immediately the symmetry group condition  to get bounds on 
$U_{a-b}$, whereas in  \cite{Lam:2014kga} the symmetries $S_a$ and $S_b$  
were used to obtain bounds on the corresponding mass matrices. 
Diagonalization of these restricted matrices and then finding the relative mixing 
should lead to the same result. 

Let us illustrate this using two examples. 
We will consider the Coxeter group $A_3$. It has three generators and     
the group structure is 
\beq
(S_1 S_3)^2 = \mathbb{I},~~~~~  (S_1 S_2)^3 = \mathbb{I}, ~~~~~ 
(S_3 S_2)^3 = \mathbb{I}. 
\label{grouprel}
\eeq

In the first example we take  $Y_{10}$ to be invariant with respect to 
$S_{10} =  S_1$  and  $Y_{126}$ with respect to  $S_{126} =   S_3$. 
From the first group relation in (\ref{grouprel}) it follows that 
$S_1$ and  $S_3$ commute. Therefore the basis can be found in which 
both $S_1$ and  $S_3$ are diagonal simultaneously. 
We can take $S_{10} =  S_1^d$  and $S_{126} =   S_3^d$,  
where $S_1^d$ and  $S_3^d$ are given in  (\ref{eq:reflect}). 

Let us underline that in this example it is the commutation of $S_1$ and  $S_3$ 
(which is a consequence of the group structure relation)  that encodes the 
information about embedding. 

As the consequence of symmetries,  
the matrices should have the following vanishing elements
\bea
\label{Y10}
\left(Y_{10}\right)_{12,13,21,31}&=&0\\
\label{Y126}
\left(Y_{126}\right)_{13,23,31,32}&=&0. 
\eea
They are diagonalized by 
\beq
\label{ambig}
U_{10}=
\bem
e^{i\alpha_{10}} & 0_{1\times 2} \cr
0_{2\times 1} & \left(U_{10}\right)_{2\times 2}
\eem\;\;\;,\;\;\;
U_{126}=
\bem
\left(U_{126}\right)_{2\times 2} & 0_{2\times 1} \cr
0_{1\times 2} & e^{i\alpha_{126}}
\eem
\eeq
Therefore
\beq
\label{U10U126}
U_{13} =   \left(U_{10}^\dagger U_{126}\right)_{13} = 0. 
\eeq

This result can be obtained immediately from our consideration 
(\ref{Uji}). 
Indeed, in this case  the generators $S_1$ and $S_3$ are involved, so we fix the element 
$U_{13}$. In this example  $p = 2$ and $k = 1$  that lead according to 
(\ref{Uji}) to $U_{13} = \cos{(\pi/2)} = 0$. 

In the second example we take again $S_{10} = S_1$ as the symmetry of $Y_{10}$ 
but $S_{126} = S_2$ as the symmetry of $Y_{126}$. Now $p = 3$ (\ref{grouprel}) 
and the generators do not commute, so they can not be diagonalized simultaneously.  
In the basis $S_{10} = S_1^d$ according to \cite{Lam:2014kga} the third element 
equals
\beq
\label{y1y1}
S_2 =\frac{1}{2}
\bem
-1   & \sqrt{2} & -1 \cr
...  & 0        &  - \sqrt{2} \cr
...  & ...      & -1
\eem. 
\eeq
This element can be represented as 
\beq  
\label{y1y2}
S_2 = U S_2^d U^\dagger,  
\eeq 
where $S_2^d = diag (-1, 1, -1)$ and,  as can be obtained explicitly from 
(\ref{y1y1}) and (\ref{y1y2}),  in  $U$   
only the second column is determined: $|U_{j2}| =  (1/2, 1/\sqrt{2}, 1/2)^T$. 
The matrix $U$ is nothing but the relative matrix which connects two diagonal bases 
for $S_i$. 
In particular,  we have $|U_{12}| = 1/2$. Again this result can be obtained immediately 
from our consideration.  Since the generators involved are 
$S_1$ and $S_2$, the 1 - 2 element  is fixed.
For $p = 3$ and $k = 1$  (or $k=2$) we have from (\ref{Uji}) 
$|U_{12}| = \cos{(\pi/3)} = 1/2$. 

Notice that in the matrix $U$ only one column is determined,  and so there is an ambiguity 
related with certain rotations. Also in the first example we could write 
the symmetry group condition as $(S_1^d S_3^d)^2 = \mathbb{I}$, that is,  $U = \mathbb{I}$ 
which is consistent with $U_{13} = 0$. Again here we have an ambiguity related to 
rotations (\ref{ambig}).

\end{document}